\documentstyle[12pt]{article}

\title{Representations of $U(1,q)$ and Constructive Quaternion Tensor
Products}
\author{by\\ S. De Leo  and  P. Rotelli\\ Dipartimento di Fisica,
Universit\`a  di Lecce\\ INFN-Sezione di Lecce}
\date{}

\begin{document}

\maketitle

\begin{abstract}
The representation theory of the group $U(1,q)$ is discussed in detail
because of its possible application in a quaternion version of the
Salam-Weinberg theory.
As a consequence, from purely group theoretical arguments we
demonstrate that the eigenvalues must be right-eigenvalues and that
the only consistent scalar products are the complex ones. We also define
an explicit quaternion tensor product which leads to a set of additional
group representations for integer ``spin''.
\end{abstract}

\pagebreak

\section*{I \ \ Introduction}

\hspace*{5mm} Quaternions have been somewhat of an enigma for Physicists since
their discovery by Hamilton  \cite{ham} in 1843.
Notwithstanding Hamilton's conviction that quaternions would soon
play a role comparable, if not greater than that of complex numbers, the
use of quaternions in Physics is very limited. Amongst the contributions
to quaternion quantum mechanics we draw attention to the fundamental
works of Finkelstein et al. \cite{fin1,fin2,fin3,fin4} (on foundations of
quaternion quantum mechanics, on quaternionic representations of compact
groups, etc),  of Horwitz and Biedenharn \cite{hor} (on quaternion quantum
mechanics, second quantization and gauge fields) and to the many stimulating
papers of Adler \cite{adl1,adl2,adl3} (on quaternion
potentials and CP violation, on quaternion field theory, etc).

Complex numbers in Physics have played a dual role, first as a technical
tool in resolving differential equations (e.g. in classical optics)
or via the theory of analytic functions for performing real integrations,
summing series etc, secondly, in a more essential way in the development
of quantum mechanics (and later field theory) characterized by complex wave
functions and for fermions by complex wave equations. With quaternions, for
the first type of application, i.e. as a means to simplify calculations, we
can quote the original work of Hamilton, but this only because of the late
development of vector algebra. Even Maxwell \cite{max} used quaternions
as a tool in his calculations. The more exciting
possibility that quaternion equations will eventually play a significant
role is synonymous, for some physicists (but not for the present authors),
with the advent of a revolution in Physics comparable to that of
quantum mechanics.

Our own particular point of view is that even if quaternions do not
simplify calculations, it would be very strange if standard quantum
mechanics did not permit a quaternion description other than in the trivial
sense that complex numbers are contained within the quaternions. In other
words, given the validity of quantum mechanics at the elementary particle
level, we predict the existence even at this level of quaternion versions
of all standard theories. One of the authors \cite{rot1,rot2}
has indeed succeeded in this
with a quaternion version of the Dirac equation. This equation, thanks to
the use of the complex scalar product, reproduces the standard results
notwithstanding the two-component nature of the wave functions due to the
existence of two-dimensional quaternion gamma matrices. The same doubling
of solutions implies that the Schr\"{o}dinger equation has automatically
two plane wave solutions corresponding to spin up and spin down. This
doubling of solutions continues even for bosonic equations. As a result,
two photonic solutions exist (one called anomalous), two scalar solutions
of the Klein-Gordon equation exist and so forth. It has also been
demonstrated \cite{del} that these new anomalous
solutions can be associated with
corresponding anomalous fields. We observe that the existence of these new
solutions implies that the use of quaternions is not without predictive power,
at least with the formalism described above.

Coherent with our point of view, it appears to us desirable to develop a
quaternion version of the Salam-Weinberg theory of electro-weak
interactions. This theory may also provide a proving ground for the
anomalous particles. For it is possible that the anomalous photon be
identified with one of the massless neutral intermediate vector fields,
prior to the creation of mass via spontaneous symmetry breaking. In other
words it has been suggested \cite{del} that the anomalous photon
could be identified with the $Z^0$.

As a preliminary to this non trivial objective one must decide upon the
appropriate quaternion version of the Glashow group $SU(2,c) \times U(1,c)$.
The $c$ here specifies {\em complex group} and implies only complex matrix
elements, i.e. standard group theory. A $q$ within a group name will imply a
quaternion group with in general quaternion matrix elements, even if this
does not exclude the appearance of purely complex or even real group
representations.
Surprisingly, the complex group $U(1,c)$ remains as such even for a
quaternion version of Salam-Weinberg. This is not difficult
to justify, but we leave this explanation to a subsequent article.
The group $SU(2,c)$ is particularly interesting, first because
this Lie group is not only the weak isospin group of Salam-Weinberg but is
very common in particle physics (spin, isospin, etc) and second
because we do indeed have an alternative choice in the quaternion unitary
group $U(1,q)$, also referred to as the symplectic group $Sp(1,q)$. It is well
known that these groups are isomorphic to $SU(2,c)$ \cite{gil}.
However, as we shall demonstrate in this paper, this does not guarantee
identical physical content.
For example, with the complex group $SU(2,c)$ all
representations are obtainable from the spinor representation with
the aid of tensor products.
This will not be the case for $U(1,q)$, and indeed the definition
of a suitable quaternion tensor product is still
of primary interest \cite{raz1,raz2,raz3}.

In the next Section we shall develop the representation theory of $U(1,q)$
in analogy with that of $SU(2,c)$ (we shall henceforth use the term ``spin''
to identify the physical observable associated with these groups). In
particular, we obtain a group
theoretical justification both for the use of the right-eigenvalue
equations and for the adoption of the complex scalar product. We also
derive a set of representations for each ``spin'' value. In Section III
we define an explicit right-complex linear tensor product in terms of
{\em quaternion} column matrices, suggested by the work of Horwitz and
Biedenharn. As a consequence, we discover additional non equivalent
(see Appendix B) matrix
representations for integer spins, characterized by the absence of
anomalous eigenstates. The physical significance of these results are
discussed in our conclusions in the final Section.

\pagebreak

\section*{II \ \  Comparison of $U(1,q)$ and $SU(2,c)$ representations}

\hspace*{5mm} The representation theory of $SU(2,c)$ is well known. We recall
in
particular the importance of the Pauli matrices which represent twice the
generators of ``spin'' 1/2. The unitary quaternion group $U(1,q)$ may be
usefully confronted with both $U(1,c)$ and $SU(2,c)$. $U(1,q)$ is in fact the
natural generalization of $U(1,c)$ to the non-commutative quaternion
numbers, defined by
\begin{equation} q=q_{0} + q_{1} i + q_{2} j + q_{3} k \end{equation}
\begin{center}
$(q_{m}\in R \ \ \ m=0, \ldots, 3)$
\end{center}
where there are three imaginary elements
$i,j,k$\  $(i^{2} = j^{2} = k^{2} = -1)$\ and
\begin{equation} i j k = -1 \: . \end{equation}
The complex numbers $C(1,i)$, with bases 1 and $i$, are a subset of the
quaternions. More precisely there are infinite, a priori, equivalent complex
planes in the four dimensional quaternion space. The above plane will
however be identified with the standard complex numbers.

The abelian group $U(1,c)$ is therefore represented by the one dimensional
general group elements
\begin{equation} g\sim e^{-i\alpha} \ \ \ \ \ \alpha \in R \end{equation}
The non abelian group $U(1,q)$ is represented at the lowest non-trivial
level by
\begin{equation} g\sim e^{-i\alpha -j\beta-k\gamma} \ \ \ \ \ \alpha,\beta,
\gamma \in R \end{equation}
If we identify the numbers  $\frac{i}{2},\frac{j}{2},\frac{k}{2}$ as the
generators of the group, these define a Lie algebra with
{\em anti-hermitian} generators $A_{m}$ ($m=1, 2, 3$) satisfying
\begin{equation} [A_{m},\  A_{n}] = \epsilon_{mnp}\ A_{p} \end{equation}
This equation is readily identified with that of the Lie algebra $su(2,c)$
if one considers as generators of the corresponding group
$-i\sigma_{m}/2$, where $\sigma_{m}$ for $m=1, 2, 3$ are the Pauli matrices.
This is the connection between $U(1,q)$ and $SU(2,c)$ which as a
consequence are isomorphic groups \cite{gil}.
We shall demonstrate in this work that the isomorphism of the abstract
groups is not automatically reflected in the corresponding matrix
representations. Indeed, we anticipate that we shall find for $U(1,q)$
two inequivalent matrix representations for the same integer spin values. We
shall also encounter generators which are formally irreducible into smaller
block structures, but which operate on a vector space which is fully
reducible.

We begin our comparison between $U(1,q)$ and $SU(2,c)$ by choosing a
{\em polarization direction}, say that corresponding to the
$\frac{i}{2}$ generator.
In $SU(2,c)$ the choice of the polarization direction leads naturally to the
choice of the corresponding eigenvectors as a basis in spin space and the
automatic diagonalization of the generator. This is of course not the case
for $U(1,q)$ since all three generators, being one dimensional, are already
``diagonal''.
We can however still seek the eigenvectors for $\frac{i}{2}$.
However, before doing so, we need to recall some facts involving
quaternions.

First note that for any quaternion operator $A$ with eigenvector $\psi$ and
eigenvalue $a$, we have, a priori, two options for the eigenvalue equation,
either the left eigenvalue equation
\begin{equation}  A\psi = a\psi \: , \end{equation}
or the right eigenvalue equation
\begin{equation}  A\psi = \psi a \: . \end{equation}
We also observe at this point that the most general quaternion operator $A$
is of the ``bared'' type $A = B\vert b$ defined by its action upon any state
vector $\psi$,
\begin{equation} A\psi \equiv B\psi b \: . \end{equation}
This complication being due to the non commutative nature of quaternions.
Obviously the modulus of $b$ may always be set to unity without loss of
generality. We further recall that with the adjoint ($^{+}$) defined as
\begin{equation}
(q_{0} + q_{1} i + q_{2} j + q_{3} k )^{+}
= q_{0} - q_{1} i - q_{2} j - q_{3} k
\end{equation}
and \begin{equation}
(q_{1}q_{2})^{+} = {q_{2}}^{+}{q_{1}}^{+}
\end{equation}
the norm of each quaternion is given by

\begin{equation}
{\vert q \vert}^{2} \: = \: q^{+}q \: > \: 0
\end{equation}
and thus each non null quaternion has an inverse
\begin{equation}
q^{-1} = \frac{q^{+}}{{\vert q \vert}^{2}} \ \ .
\end{equation}
Now if, as in our case, $A^{+} = -A$ then, depending upon the use of the
left or right eigenvalue equation, we will have, either
\begin{equation} \psi^{+}A\psi = \psi^{+} a\psi \end{equation}
or
\begin{equation} \psi^{+}A\psi = \psi^{+}\psi a \: . \end{equation}
In either case by taking adjoints we readily demonstrate that $a$ does not
have a real part,
\begin{equation}  a^{+} = -a \end{equation}
(notice that in the latter case we need the fact that $\psi^{+}\psi$
is real and hence commutes with any $a$).

At this point we run the risk of having infinite eigenvectors for
(spin) $s=\frac{1}{2}$.
To impose only a finite number of solutions to our eigenvalue
equations, we select a preferential complex plane, that based upon the
units 1 and $i$. We then require that the eigenvalues be proportional to
the $i$ unit. This accords well with the eigenvalues of
$-i\sigma_{m}/2$, the $SU(2,c)$ counterpart. Now our two choices
of eigenvalue equation may be written explicitly as
\begin{equation} \frac{i}{2}\psi_{L} = (-i\lambda)\psi_{L}
\end{equation}
or \begin{equation} \label{a}
\frac{i}{2}\psi_{R} = \psi_{R}(-i\lambda)
\end{equation}
with $\lambda \in R $.
It is easy to solve these equations. The left-eigenvalue equation is
right-quaternion linear (if $\psi$ is a solution, so is $\psi q $ with $q$
any quaternion) but it has only one eigenvalue compared to the two
($\lambda = \pm\frac{1}{2}$) of $SU(2,c)$. Only the right-eigenvalue
equation (eq.(\ref{a})) yields the two desired
eigenvalues, corresponding respectively to the eigenvectors $\psi=jz$ and
$\psi=z^{\prime}$, where $z$ and $z^{\prime}$ are arbitrary $C(1,i)$ complex
numbers. These solutions are characterized by being only right-complex
linear. Ignoring for the moment the right complex phases ($z$,$z'$) we
can summarize our results by saying that to conform with standard
$SU(2,c)$ results for spin $\frac{1}{2}$ we are obliged to choose
the right-eigenvalue equation and as a consequence obtained the solutions,
\begin{equation}
\psi_{+}=j \Leftrightarrow \lambda = \frac{1}{2} \end{equation}
\begin{equation}
\psi_{-}=1 \Leftrightarrow \lambda = -\frac{1}{2} \end{equation}
The choice of the generator is, as expected, not of any significance in all
this. If we had ``diagonalized'' the $\frac{k}{2}$ generator, we would not have
found any left-eigenvectors at all, but exactly two right-eigenvectors
$(1\pm j)z$ with the same $\lambda$ eigenvalues $\pm \frac{1}{2}$ found
previously.

We have not yet finished with spin $\frac{1}{2}$, in $SU(2,c)$ the
two eigenvectors are automatically orthogonal. This is not the case for
$U(1,q)$. Indeed in order to impose orthogonality we are obliged to adopt
what is known as the ``complex scalar product''. The quaternion scalar product
is defined as a straightforward generalization of the standard complex
counterpart. Thus for {\em simple numbers} this is defined by
\begin{equation}
\langle f\vert g\rangle\equiv f^{+} g\ .\end{equation}
The complex scalar product with quaternions, is simply the projection of the
above upon the privileged complex plane, i.e. it is defined as follows
\begin{equation}
(f\vert g)\equiv \frac{1}{2} [\langle f\vert g\rangle - i \langle
f\vert g\rangle i]\end{equation}
In this way 1 and $j$ are complex orthogonal, as occurs for the
corresponding eigenvectors in $SU(2,c)$.

It is interesting to recall that
the complex scalar product was first proposed by Horwitz and Biedenharn in
1984 \cite{hor} for quaternion quantum mechanics as a means of performing
quaternion tensor products and defining a suitable Fock space.
Subsequently, it was rederived from a quaternion Dirac equation by
the natural requirement that the (non-conventional) momentum operator
be hermitian. The present derivation
has the merit of being a consequence of group theoretical arguments,
independent of any physical dynamical equation of motion. It must however be
admitted that Adler has in recent years become a fervent advocate of the use
of a quaternion scalar product. It would not then be possible to reproduce
with $U(1,q)$ the properties of $SU(2,c)$. Indeed we suspect that only the
{\em complex groups} could be employed with this purer quaternion
hypothesis, if one wishes to preserve standard group theory in Physics.
Paradoxically this contradicts the a priori rejection of a preferential
complex plane.

We also observe that with the introduction of the complex scalar
product it is possible to pass from the intrinsic anti-hermitian nature of
the generators of $U(1,q)$ to an equivalent (complex) hermitian set $J_m$
defined by
\begin{equation}
\frac{i}{2} \: ,\: \frac{j}{2} \: , \frac{k}{2} \: \Leftrightarrow
\frac{i}{2}\vert i \: , \: \frac{j}{2}\vert i \: , \: \frac{k}{2}\vert i
\end{equation}
which satisfy the Lie algebra for the {\em generalized angular momentum} if
this algebra is redefined as:
\begin{equation}
[ J_{m},\ J_{n}] = \epsilon_{mnp} \, J_{p}\vert i \end{equation}
{}From this algebra, following the standard method of derivation (with some
care in the position of the $i$ factors) it is possible to derive the
well known angular momentum spectrum. We emphasize that this can only
be done after adoption of the complex scalar product.

As an aid in determining the quaternion generators for higher spin values,
we observe that the $-i\sigma_{m}/2$ of $SU(2,c)$  are reducible with
quaternions (Appendix A) to a diagonal form with
our generators $A_{m}$ as elements (this
is not true for $\pm\sigma_{m}/2$ {\em without} the $i$ factor). Thus
we expect that all semi-integer spin value representations can be derived by
reduction from the corresponding $SU(2,c)$ generators, yielding matrices
with exactly half the dimensions.

For integer spin values the complex representations derivable from those
of $SU(2,c)$ are not reducible
even with quaternions, but we shall justify this claim only in Appendix B.
As a consequence, for integer spins, we encounter ``anomalous''
eigenvector solutions obtained from any standard complex eigenvector by
simply multiplying from the right by $j$. For example, for spin 1 the
complete set of eigenvectors are

\[ \left(\begin{array}{c} 1\\0\\0\end{array}\right),
\left(\begin{array}{c} 0\\ 1\\ 0\end{array}\right),
\left(\begin {array}{c} 0\\ 0\\ 1\end{array}\right);
\left(\begin {array}{c} 1\\ 0\\ 0\end{array}\right)j \, ,
\left(\begin{array}{c} 0\\ 1\\ 0\end{array}\right)j \, ,
\left(\begin {array}{c} 0\\ 0\\ 1\end{array}\right)j\]
and furthermore the two sets of states divided by the semi-colon above are
manifestly invariant subspaces (recall that the generators and hence group
elements are complex for this case). Thus we encounter the situation,
anticipated in the introduction, of non reducible matrix representations
for the generators of the group $U(1,q)$ notwithstanding a completely reducible
eigenvector space.

Let us list the first few representations found,\\
$s=0$\ \ :\ \   $A_{1}=A_{2}=A_{3}=0$\\
$s=\frac{1}{2}$\ \ :\ \  $A_{1}=\frac{i}{2}$, $A_{2}=\frac{j}{2}$,
$A_{3}=\frac{k}{2}$\\
$s=1$\ \ :\ \ Standard complex representation\\
$s=\frac{3}{2}$\ \ :\ \  \[ A_{1}=\left( \begin{array}{cc}
\frac{3i}{2} & 0 \\
0 & \frac{i}{2} \end{array}\right)\ ,
\: A_{2} = \left(\begin{array}{cc}
0 & \frac{\sqrt{3}}{2}\\
-\frac{\sqrt{3}}{2} & j \end{array}\right)\ ,
\: A_{3}= \left(\begin{array}{cc}
0 & \frac{\sqrt{3}\,i}{2}\\
\frac{\sqrt{3}i}{2}  & k \end{array}\right) \]
We can tabulate the general half integer spins representation as
follows
\vspace*{2mm}
\begin{eqnarray}
A_{1} & = & \left(\begin{array}{cccccc}
si & 0 & 0 &\bullet & 0 & 0\\
0 & (s-1)i & 0 & \bullet & 0 & 0\\
0 & 0 & (s-2)i & \bullet & 0 & 0\\
\bullet & \bullet & \bullet & \bullet & 0 & 0\\
0 & 0 & 0 & 0 & \frac{3}{2} i & 0\\
0 & 0 & 0 & 0 & 0 & \frac{1}{2} i\end{array}\right) \nonumber \\ \nonumber \\
A_{2} & = & \left(\begin{array}{cccccc}
0 & a & 0 & \bullet & 0 & 0\\
-a & 0 & b & \bullet & 0 & 0\\
0 & -b & 0 & \bullet & 0 & 0\\
\bullet & \bullet & \bullet & \bullet & v & 0 \\
0 & 0 & 0 & -v & 0 & z\\
0 & 0 & 0 & 0 & -z & \alpha j\end{array}\right) \\ \nonumber \\
A_{3} & = & \left(\begin{array}{cccccc}
0 & ai & 0 & \bullet & 0 & 0\\
ai & 0 & bi & \bullet & 0 & 0\\
0 & bi & 0 & \bullet & 0 & 0\\
\bullet & \bullet & \bullet & \bullet & vi & 0\\
0 & 0 & 0 & vi & 0 & zi\\
0 & 0 & 0 & 0 & zi & \alpha k\end{array}\right) \nonumber
\end{eqnarray}
The conditions
\begin{eqnarray}
A^{2} & \equiv & A^{2}_{1} + A^{2}_{2} +
A^{2}_{3} = -s(s+1) \nonumber \\
{\epsilon}_{mnp}A_{p} & =  & [A_{m},A_{n}] \\
A^{+}_{m} & = & -A_{m} \nonumber
\end{eqnarray}
then determine the corresponding matrix elements
\begin{eqnarray}
\begin{array}{lllclll}
a & = & \sqrt{\frac{s}{2}} & & d & = & \sqrt{\frac{4s-6}{2}}\\
b & = & \sqrt{\frac{2s-1}{2}} & & e & = & \sqrt{\frac{5s-10}{2}}\\
c & = & \sqrt{\frac{3s-3}{2}} & & & {\rm etc.} & \end{array}
\end{eqnarray}
For the first few half-integer spins these results are collected in Table 1

\begin{center}

\begin{tabular}{||r|c|c|c|c|c|c|r||}
\hline\hline
spin & a & b & c & d & e & f & $\alpha$\\
\hline\hline
$\frac{1}{2}$ & & & & & & & $\frac{1}{2}$\\
\hline
$\frac{3}{2}$ & $\sqrt{\frac{s}{2}}$ & & & & & & 1\\
\hline
$\frac{5}{2}$ & '' & $\sqrt{\frac{2s-1}{2}}$ & & & & & $\frac{3}{2}$\\
\hline
$\frac{7}{2}$ & '' & '' & $\sqrt{\frac{3s-3}{2}}$ & & & & 2\\
\hline
$\frac{9}{2}$ & '' & '' & '' & $\sqrt{\frac{4s-6}{2}}$ & & & $\frac{5}{2}$\\
\hline
$\frac{11}{2}$ & '' & '' & '' & '' & $\sqrt{\frac{5s-10}{2}}$ &  & 3\\
\hline
$\frac{13}{2}$ & '' & '' & '' & '' & '' & $\sqrt{\frac{6s-15}{2}}$ &
$\frac{7}{2}$\\
\hline\hline
\end{tabular}
\end{center}
\begin{center} {\bf Table 1}: Half-integer spin\\
matrix element coefficients
\end{center}

In conclusion we have found a spin spectrum for $U(1,q)$ in conformity with
those of $SU(2,c)$, except that the dimension of the semi-integer
representations are halved and the integer spin complex representations
exhibit anomalous solutions. As we shall see in the next section, not all
is satisfactory and indeed the above set of representations are not complete.

\pagebreak

\section*{III \ \  Constructive Quaternion Tensor product and New Integer
Representations}

\hspace*{5mm} There is a difficulty with the representations found in the
previous section.
The multiplicity of states for a given integer spin are double
those of $SU(2,c)$. Apart from the question of the physical interpretation
of the anomalous states, we have a problem with the multiplicities of the
yet undefined tensor product. For example, consider the spin (tensor) product
and decomposition
\begin{equation}
\frac{1}{2} \: \otimes \: \frac{1}{2} \: =  \: 1 \: \oplus \: 0 ,
\end{equation}
which in $SU(2,c)$ corresponds to the multiplicity count $2\times2=3+1$.
For $U(1,q)$ the multiplicity count fails because the number of states with
spin one is six and those with spin zero is two. Apart from the drastic
choice of abandoning the tensor product, the solution to this incongruence is
potentially very interesting. Perhaps we must reinterpret the significance
of the tensor products, or admit that the anomalous solutions are spurious
and can be avoided in some way (other than abandoning the complex scalar
product). In any case another difference between the two groups must be
admitted. The solution we present below is based upon the definition of an
explicit (constructive) tensor product which, even if not the final word
upon quaternion tensor products, has the undoubted merit of bringing to the
light a new set of integer spin representations. We shall see that the
above failure in the multiplicity counts is due to the fact that {\em none}
of the states of integer spin so far listed are ``created'' in the tensor
product. The new integer spin states will have the {\em correct}
multiplicities.

The difficulty in defining quaternion tensor products lies in the non
commutative property of quaternions. For one component functions in quantum
mechanics we simply multiply the functions to obtain the tensor
product. This is possible because the product of two complex functions of
independent variables continue to satisfy their individual dynamical
equations of motion. With matrices the standard rules for tensor products
maintain the same feature thanks to the commutativity of complex numbers.
All this is lost with quaternion functions. For our particular case we
would also be faced with the additional problem that the algebraic product
of two quaternion numbers (spin $\frac{1}{2}$ generators) being itself a
quaternion number does not have the correct dimensions to represent any
higher spin state.

As we have anticipated in the introduction, it was Horwitz
and Biedenharn \cite{hor}
who first proposed the complex scalar product together with a corresponding
quaternion tensor product which satisfies the property of being only
right-complex linear (and not quaternion right linear). These authors write
each quaternion $q$ in {\em symplectic} form
\begin{equation} q = z + j \,   z'  \end{equation}
and thus represent $q$ as an ordered pair of complex numbers, which we write
as a column matrix rather than in row form as originally done,
\begin{eqnarray}
q & \sim & \left( \begin{array}{c} z \\ z^{\prime} \end{array} \right)
\end{eqnarray}
The tensor product $q_{1} \otimes q_{2}$
of Horwitz and Biedenharn is then defined to be
\begin{eqnarray}
q_{1} \: \otimes \: q_{2} & \sim & \left( \begin{array}{c} z_{1} \\
z_{1}^{\prime}  \end{array} \right) \: \otimes \: \left( \begin{array}{c}
z_{2}\\
z_{2}^{\prime}   \end{array}  \right)  \nonumber \\
   & = & \left( \begin{array}{c}
z_{1}z_{2}\\ z_{1}z_{2}'\\
z_{1}^{\prime}z_{2}\\ z_{1}^{\prime}z_{2}^{\prime}
\end{array} \right)
\end{eqnarray}
This formalism together with the standard rules for complex matrix scalar
products is such that,
\begin{equation}
{\vert q \vert}^{2} = {\vert z \vert}^{2} + {\vert z' \vert}^{2}
\end{equation}
\begin{equation}
(q_{1}\otimes q_{2}\vert \, q_{3}\otimes q_{4} ) = (q_{1}\vert q_{3} )
(q_{2}\vert q_{4} ) \end{equation}
where we recall that $ ( \: \vert\: ) $ is the {\em complex} scalar product
between quaternions.

This formalism has a defect in that beyond the first level (single
particle) one loses sight of any possible quaternion structure in the
tensor products. The symplectic formalism confirms however our previous
results in that
\begin{eqnarray}
1 \: \sim \: \left( \begin{array}{c} 1\\
0 \end{array} \right) & and & j \: \sim \: \left( \begin{array}{c} 0\\
1 \end{array} \right)
\end{eqnarray}
so that the analogy between $U(1,q)$ and $SU(2,c)$ at the spin
$\frac{1}{2}$
level is manifest. We now suggest the following modification in the above
formalism. We define a two component quaternion column matrix as the result
of a tensor product between two quaternion numbers (or functions),
\begin{eqnarray}
q_{1} \: \otimes \: q_{2} & = & \left( \begin{array}{c}
q_{1}z_{2}\\ q_{1}z_{2}' \end{array} \right)
\end{eqnarray}
where $z_{2}$ and $z_{2}^{\prime}$ are as before the symplectic parts of
$q_{2}$. This definition is still right-complex linear both for $q_{1}$ and
$q_{2}$. The complex scalar products between tensor products continue to
satisfy the same decomposition property as above, as the following equations
demonstrate,
\begin{eqnarray}
(q_{1}\otimes q_{2}\vert \, q_{3}\otimes q_{4} ) & \equiv & \left( (z_{2}^{*}
q_{1}^{+} \: \: z_{2}^{\prime \, *} q_{1}^{+} ) \left( \begin{array}{c}
q_{3}z_{4}\\ q_{3}z_{4}^{\prime} \end{array} \right) \right)_{C} \nonumber \\
  & = &  {( z_{2}^{*}q_{1}^{+}q_{3}z_{4} + {z_{2}'}^{*}q_{1}^{+}q_{3}
z_{4}^{\prime} )}_{C} \nonumber \\
  & = &  z{_2}^{*}{(q_{1}^{+}q_{3})}_{C}z_{4}+{z_{2}'}^{*}{(
q_{1}^{+}q_{3})}_{C}z_{4}' \nonumber \\
  & = & {(q_{1}^{+}q_{3})}_{C} (z_{2}^{*}z_{4}+z_{2}^{\prime \, *}
z_{4}^{\prime} ) \nonumber \\
  & = & {(q_{1}^{+}q_{3})}_{C}{(q_{2}^{+}q_{4})}_{C} \nonumber \\
  & = & (q_{1}\vert q_{3})(q_{2}\vert q_{4})
\end{eqnarray}
as before. In the above derivation $(q)_{C}$ stands for the complex
$C(1,i)$
projection of the quaternion $q$. Our formalism maintains a quaternion
structure for the tensor products and is readily generalizable, e.g.
\begin{eqnarray}
q_{1} \: \otimes \: q_{2} \: \otimes \: q_{3} & = & \left( \begin{array}{c}
q_{1}z_{2}\\ q_{1}z_{2}^{\prime}
\end{array} \right) \: \otimes \: q_{3} \nonumber \\
  & \equiv & q_{1} \: \otimes \: \left( \begin{array}{c}
q_{2}z_{3}\\ q_{2}z_{3}^{\prime} \end{array} \right) \nonumber \\
  & = & \left( \begin{array}{c}
q_{1}z_{2}z_{3}\\ q_{1}z_{2}z_{3}^{\prime}\\
q_{1}z_{2}^{\prime}z_{3}\\ q_{1}z_{2}^{\prime}z_{3}^{\prime}
\end{array} \right)
\end{eqnarray}

The main advantage for us in this formalism will be the extraction of a set
of generators for $s=1\oplus 0$ obtainable directly from those of spin
$\frac{1}{2}$ and automatically consistent
with the multiplicity rules. It is
immediately obvious from the matrix dimensions of the tensor product
$q_{1}\otimes q_{2}$ that the generators in question will necessarily be of
dimension two (if not reducible) in contrast with those of spin 1 $\oplus$
spin 0 discussed in the previous section, which have dimension four (3+1).
Indeed after a straightforward calculation we find the (unseparated)
generators of spin $1\oplus 0$ to be:
\begin{eqnarray} \label{b}
\frac{1}{2} \left( \begin{array}{cc}
i + 1\vert i &  0 \\ 0 & i -1\vert i \end{array} \right) ,
\: \frac{1}{2} \left( \begin{array}{cc}
j & 1\vert i \\ 1\vert i & j \end{array} \right) ,
\: \frac{1}{2} \left( \begin{array}{cc}
k & -1\\ 1 & k \end{array} \right)
\end{eqnarray}
Note that $[a\vert b]\times[c\vert d]=ac\vert bd$ and that all simple
quaternion numbers $q$ are formally $q\vert 1$. The eigenvectors of the
first generator above are:
\begin{eqnarray}
s \: = \: 1 \: : &  & \left( \begin{array}{c} 0\\ j \end{array} \right),
\: \frac{1}{\sqrt{2}} \left( \begin{array}{c} j\\
i \end{array} \right),
\: \left( \begin{array}{c} 1\\ 0 \end{array} \right) \nonumber \\
s \: = \: 0  \: :&  & \frac{1}{\sqrt{2}} \left( \begin{array}{c}
-j\\ i \end{array} \right)
\end{eqnarray}
It is to be noted that one does not obtain a second set of complex
orthogonal eigenvectors to those listed above by multiplying from the right
by $j$, as occurs with the anomalous solutions. This is due to the
somewhat surprising fact that here $A^{2}$  is neither diagonal nor
proportional to the identity matrix. Indeed the expression for $A^{2}$ is
\begin{eqnarray}
A^{2} & = & -\frac{1}{2} \left( \begin{array}{cc}
3 - i\vert i & k - j\vert i \\
-k - j\vert i & 3 + i\vert i  \end{array} \right)
\end{eqnarray}
and the barred nature of its matrix elements means that right
multiplication by $j$ will not leave the eigenvalue of $A^{2}$ unaltered.
The four solutions are simply mixed under right multiplication by $j$. We
also observe that the same generators are valid both for spin 1 and spin 0
and they cannot be further reduced.

If this analysis is extended to $\frac{1}{2}\otimes\frac{1}{2}\otimes
\frac{1}{2}$ we obtain the generators for $\frac{3}{2}\oplus
\frac{1}{2}\oplus \frac{1}{2}$
\vspace*{2mm}
\begin{eqnarray}
A_{1} & = & \frac{1}{2} \left( \begin{array}{cccc}
2\vert i + i & 0 & 0 & 0\\ 0 & -2\vert i +i & 0 & 0\\
0 & 0 & i & 0\\
0 & 0 & 0 & i \end{array} \right) \nonumber \\ \nonumber \\
A_{2} & = & \frac{1}{2} \left( \begin{array}{cccc}
j & 0 & 1\vert i & 1\vert i\\ 0 & j & 1\vert i & 1\vert i\\
1\vert i & 1\vert i & j & 0\\
1\vert i & 1\vert i & 0 & j \end{array} \right) \\ \nonumber \\
A_{3} & = & \frac{1}{2} \left( \begin{array}{cccc}
k & 0 & -1 & -1\\ 0 & k & 1 & 1\\ 1 & -1 & k & 0\\
1 & -1 & 0 & k \end{array} \right) \nonumber
\end{eqnarray}

\pagebreak

with eigenvectors (in decreasing order of $s_{x}$)
\begin{eqnarray}
s \: = \: \frac{3}{2} \: : &  & \left( \begin{array}{c}
0 \\ j \\ 0 \\ 0 \end{array} \right),
\:  \sqrt{\frac{1}{3}} \left( \begin{array}{c}
0 \\ i \\ j \\ j \end{array} \right),
\: \sqrt{\frac{1}{3}} \left( \begin{array}{c}
j \\ 0 \\ i \\ i \end{array} \right),
\: \left( \begin{array}{c}
1 \\ 0 \\ 0 \\ 0 \end{array} \right) \nonumber \\
s \: = \: \frac{1}{2} \: : &  & \sqrt{\frac{2}{3}} \left( \begin{array}{c}
0 \\ -\frac{i}{2} \\ j \\ -\frac{j}{2} \end{array} \right),
\: \: \sqrt{\frac{2}{3}} \left( \begin{array}{c}
-\frac{j}{2}\\ 0 \\ -\frac{i}{2} \\ i \end{array} \right) \\
s \: = \: \frac{1}{2} \: :&  & \sqrt{\frac{1}{2}} \left( \begin{array}{c}
0 \\ i \\ 0 \\ -j \end{array} \right),
\: \: \sqrt{\frac{1}{2}} \left( \begin{array}{c}
-j \\ 0 \\ i \\ 0 \end{array} \right) \nonumber
\end{eqnarray}
The significant fact here is that there exists an invertible quaternion
matrix $S$ which transforms the above results for
$\frac{1}{2}\otimes \frac{1}{2}\otimes\frac{1}{2}$
into those of the previous section.\\
Explicitly,
\[S \: = \: \left( \begin{array}{cccc}
\frac{1}{2} (1-i\vert i) & \frac{1}{2} (1+i\vert i) & 0 & 0\\
\frac{1}{2\sqrt{3}} k (i-1\vert i) & \frac{1}{2\sqrt{3}} k
(i+1\vert i) & \frac{1}{\sqrt{3}}\vert i & \frac{1}{\sqrt{3}}\vert i \\
\frac{1}{2\sqrt{2}} k (1+i\vert i) & \frac{1}{2\sqrt{2}} k
(1-i\vert i) & \frac{1}{2\sqrt{2}} (1-i\vert i) & \frac{1}{2\sqrt{3}}
(-1-i\vert i)\\
-\frac{k\sqrt{2}}{4\sqrt{3}} (1+i\vert i) & -\frac{k\sqrt{2}}{4\sqrt{3}}
(1-i\vert i) & \frac{\sqrt{2}}{4\sqrt{3}}
(3+i\vert i) & \frac{\sqrt{2}}{4\sqrt{3}} (-3+i\vert i)
\end{array} \right) \]
with $S^{-1}=S^{+}$.

Thus nothing new is found as far as the ``fermionic'' modes are concerned.
On the other hand we have a new and alternative set of ``bosonic'' states and
corresponding generators (distinguished by the barred matrix elements) with
the correct multiplicity. This obliges us to rediscuss the significance of
the integer representations with anomalous solutions and this will be done
in the final Section.

\pagebreak

\section*{IV \ \   Conclusions}

\hspace*{5mm} The study of the representations of $U(1,q)$ has yielded results
of interest, some of which are indeed surprising. We summarize the more
important results found in the previous Sections.

\vspace*{4mm}

1) The groups $U(1,q)$ and $SU(2,c)$ are isomorphic but the corresponding
representation structures are not.

\vspace*{4mm}

2) In order to reproduce the basic features of $SU(2,c)$, e.g. the number
of eigenvectors for ``spin'' $\frac{1}{2}$, one must adopt the
right-eigenvalue equation within $U(1,q)$.

\vspace*{4mm}

3) Even then the two solutions for spin $\frac{1}{2}$ are not orthogonal
unless one further adopts the complex scalar product.

\vspace*{4mm}

4) The quaternion representations for ``fermionic'' states have half the
dimensions of the equivalent $SU(2,c)$ cases. The total number of states
remains the same because of the doubling due to the complex scalar product.

\vspace*{4mm}

5) The known integer representations of $SU(2,c)$ are also representations
of $U(1,q)$ characterized by the existence of anomalous solutions.

\vspace*{4mm}

6) If these integer representations were unique, we would have difficulty
in the multiplicity count for tensor product states. This can be claimed
even in the absence of a specific definition of the tensor product.

\vspace*{4mm}

7) Thanks to an explicit tensor product, suggested by the work of Horwitz
and Biedenharn \cite{hor}, a new set of integer spin representations have been
found.
These automatically satisfy the multiplicity counts. Nothing new in the
fermionic sector appears.

\vspace*{4mm}

8) These new representations have lower dimensions than those previously
described and are without anomalous solutions. They are also characterized
by the presence of barred operators in some matrix elements. As with the
original set of integer spin representations, these new generator (and
hence group) representations are not reducible into smaller block form but
operate upon a reducible vector space (spin 1 plus 0 in the example treated).

\vspace*{4mm}

The use of right eigenvalue equations and the need for a complex scalar
product, which thus breaks the $i$,$j$,$k$ symmetry of quaternions, are well
known results, even if, as we have already noted the latter is not
universally accepted. However, the derivation in this work is new and
independent of any specific physical input such as the use of a particular
quaternion version of the Dirac equation. Indeed our only objective is to
reproduce the maximum of agreement between the representation theories of
$U(1,q)$ and $SU(2,c)$. This has been achieved sufficiently to suggest
that with quaternions the Glashow group could indeed be
$U(1,q)\times U(1,c)$. The question that remains to be answered is what
physical significance are we to ascribe to the alternative set of integer
spin representations i.e. those with the anomalous solutions. We recall
that these appear automatically in the solution of all known, integer
spin equations such as the Maxwell equation.

We first give a heuristic argument why the two sets of integer
spin representations are not equivalent. Even the most general unitary
similarity transformation ($S^{+}=S^{-1}$) cannot alter the
complex orthogonality of the standard and anomalous solutions. This is
because the adjoint operator is only definable for matrices with at most
$q\vert i$ elements (no $\vert j$ or $\vert k$ allowed). On the other
hand we have seen that with the alternative tensor product representations
there are no anomalous solutions. Furthermore we observe that the number
of eigenvectors are different for the  alternative sets of integer
representations (but see Appendix B).

The physical interpretation at this point
can only be speculative. We shall consider the integer states created in the
tensor products as ``composite'' (quark ?) states. The alternative series will
be, in contrast, considered elementary or fundamental. The solutions of the
Maxwell and Klein-Gordon equations etc would then correspond to fundamental
particles as opposed to composite states (such as perhaps the rho particle).
The acid test of such an interpretation would be the existence of new
integer (composite) particle equations, probably of quaternion form, i.e.
non existent within standard complex quantum mechanics. Needless to say,
the identification of such a new class of quaternion integer spin
equations would be extremely
interesting, independent of their physical success. Anticipating our work
upon the quaternion version of the Salam-Weinberg model we therefore expect
that the intermediate vector bosons be elementary solutions of the standard
spin-one equations. As for the Higgs bosons, we have two alternative
choices depending upon whether we consider them elementary or composite. But
to treat this latter possibility we must first discover the equation
corresponding to a {\em composite} spin zero state.

\pagebreak

\section*{Appendix A}

\hspace*{5mm} Here we derive an explicit quaternion similarity transformation
which reduces completely the generators $-i\sigma_{m}/2$. We have already
noted in the text that these anti-hermitian generators satisfy the same
algebra as $Q_{m}/2$ where $(Q_{1},Q_{2},Q_{3})=(i,j,k)$.
Consequently, we expect that there exists an invertible matrix $S$ such
that
$$ S (-i\sigma_{m}) S^{-1} \: = \: \left( \begin{array}{cc}
Q_{m} & 0 \\ 0 & Q_{l} \end{array} \right) \eqno(A.1) $$
(where, $m, l = 1, 2, 3$). It is not, a priori, necessary that $m=l$
in the above formula. However any $Q_{l}$ can always be transformed
by another similarity transformation into $Q_{m}$.
In, let $T$ be such a transformation, i.e.,
\begin{eqnarray*}
T i T^{-1} & = & j \\
T j T^{-1} & = & k \\
T k T^{-1} & = & i
\end{eqnarray*}
then it is straightforward to show that, up to an arbitrary constant,
\[ T \: \sim \: \frac{1}{2} (1+i+j+k). \]
Thus without loss of generality we can search for an $S$ such that,

$$ S (-i\sigma_{m}) S^{-1} \: = \: Q_{m} \left( \begin{array}{cc}
1 & 0 \\ 0 & 1 \end{array} \right) \: .  \eqno(A.2) $$

\vspace*{2mm}

As an example we derive this $S$ explicitly. Let
\[ S \: = \: \left( \begin{array}{cc}
a & b \\ c & d \end{array} \right) \]
where, in principle, $a,b,c$ and $d$ may be barred quaternion numbers. Then
eq.(A.2) implies that
$$ \left( \begin{array}{cc} a & b\\
c & d \end{array} \right) (-i\sigma_{m})  \: =  \:  Q_{m} \left(
\begin{array}{cc} a & b\\
c & d \end{array} \right) \eqno(A.3) $$
Now ,
\[ \{ -i\sigma_{m} \} \: : \: \left( \begin{array}{cc} 0 & -i\\
-i & 0 \end{array} \right), \: \left( \begin{array}{cc} 0 & -1\\
 1 & 0 \end{array} \right), \: \left( \begin{array}{cc} -i & 0\\
 0 & i \end{array} \right). \]
Only two of these matrices need to be inserted in eq.(A.3) since the validity
of the algebra will then guarantee the result for the third. Using
$-i\sigma_{1}$ we find
\[ i \left( \begin{array}{cc} a & b\\ c & d \end{array} \right) \:
= \: - \left( \begin{array}{cc} b & a\\ d & c \end{array} \right) i \]
i.e., $iai=b$ and $ici=d$.
{}From $-i\sigma_{2}$ we obtain
\[ \left( \begin{array}{cc} b & -a\\ d & -c \end{array} \right) \:
= \: j \left( \begin{array}{cc} a & b\\ c & d \end{array} \right) \]
i.e., $b=ja$ and $d=jc$.
These results limit the form of $a,b,c$ and $d$ to

\begin{eqnarray*}
a & = & a_{0}(1+j) + a_{1}(i-k) \\
b & = & - a_{0}(1-j) - a_{1}(i+k) \\
c & = & c_{0}(1+j) + c_{1}(i-k) \\
d & = & - c_{0}(1-j) - c_{1}(i+k)
\end{eqnarray*}
\[ (a_{0}, a_{1}, c_{0}, c_{1} \in R) \]

We now  require that $S$ be a unitary operator $SS^{+}=S^{+}S=1$. The off
diagonal elements of this condition yields,
\[ a_{0}c_{0} + a_{1}c_{1} = 0 \]
while the diagonal elements yield,
\[ {\vert a\vert}^{2} + {\vert b\vert}^{2} = 4 a_{0}^{2} + 4 a_{1}^{2} = 1 \]
\[ {\vert c\vert}^{2} + {\vert d\vert}^{2} = 4 c_{0}^{2} + 4 c_{1}^{2} = 1 \]
An acceptable solution to these equations is $c_{0}=a_{1}=0$ and
$a_{0}=c_{1}=\frac{1}{2}$, whence
$$ S \: = \: \frac{1}{2} \left( \begin{array}{cc}
1+j & j-1\\ i-k & -(i+k) \end{array} \right) \eqno(A.4) $$

A similar reduction of $\sigma_{m}/2$ without the $i$ factor
{\em cannot} be achieved. The simplest way to demonstrate this is to recall
that a similarity transformation leaves unaltered any algebra satisfied by
the transformed quantities. Now since the $\sigma_{m}/2$ are
hermitian the eventual reduction should exhibit diagonal matrix elements of
the $J_{m}$ type ($\frac{i}{2}\vert i$ etc). However, $\sigma_{m}/2$
and $J_{m}$ do not satisfy the same algebra. Indeed,
\[ \left[ \, \frac{\sigma_{m}}{2}, \frac{\sigma_{n}}{2} \, \right] \: = \:
i \, \epsilon_{mnp} \, \frac{\sigma_{p}}{2} \]
while,
\[ [ \, J_{m}, J_{n} \,] \: = \: \epsilon_{mnp} \, J_{p}\vert i \]
the diverse positions of the $i$ factors is essential and excludes the
possibility of reduction.

\pagebreak

\section*{Appendix B}

\hspace*{5mm} In this appendix we consider the matrix
representations for integer spin values, and in particular the spin 1 and
spin 0 cases. For our convenience we will call
with the adjective {\em old} the standard complex matrix
representations (characterized by the presence of {\em anomalous}
solutions) and with {\em new} the additional quaternionic matrix
representations (the one for spin 1 $\oplus$ 0 is listed in the third
section, eq.(\ref{b})).
First we note that the {\em old} matrix representation for
spin 1 alone is certainly irreducible. This follows from the fact that a
reduction of a $3\times3$ matrix necessarily yields a one dimensional
representation of the algebra, and it is readily demonstrated that there is
no such representation for spin 1. In fact only spin 0 and spin
$\frac{1}{2}$ have one dimensional representations.

Thus it seems that the only hope to reduce to blocks of smaller
dimensions the {\em old} representations is to compare the {\em old}
complex $4\times4$ matrix, which corresponds to spin 1 $\oplus$ 0,
with a quaternionic matrix of type
$$ F_{m} \: = \: \left( \begin{array}{cc}
B_{m}^{new} & 0 \\ \bullet & C_{m}^{new} \end{array} \right) \eqno(B.1) $$
where $B_{m}^{new}, \bullet, C_{m}^{new}$ are quaternion $2\times2$
matrices.

We are thus searching for a $4\times4$ quaternion matrix $S$ which operates
upon the combined spin 1 $\oplus$ 0 space and such that,
\[ S A_{m}^{old} S^{-1} \: = \: F_{m} \]
Whence, $B_{m}^{new}$ and $C_{m}^{new}$ satisfy the same algebra as
$A_{m}^{old}$. It is sufficient to concentrate our attention upon
$A^{2 \, old}$ and use the fact that it is diagonal. Now,
$$ S A^{2 \, old} S^{-1} \: = \: F^{2} \eqno(B.2) $$
where $F^{2}=F_{1}^{2}+F_{2}^{2}+F_{3}^{2}$ etc.
This equation can be write as follows since S commutes with the identity,
$$ -2 \left( \begin{array}{cc}
1 & 0\\ 0 & 1 \end{array} \right) \: + \: 2  S \left( \begin{array}{cc}
0 & 0\\ 0 & M \end{array} \right) S^{-1} \: = \: \left( \begin{array}{cc}
B^{2 \, new} & 0\\ \bullet & C^{2 \, new} \end{array} \right) \eqno(B.3) $$
with
\[ M \: = \: \left( \begin{array}{cc}
0 & 0\\ 0 & 1 \end{array} \right) \]
Now using for $S$ and $S^{-1}$ the {\em partial} expressions
\[ S \: = \: \left( \begin{array}{cccc}
\bullet & \bullet & \bullet & a \\
\bullet & \bullet & \bullet & b \\
\bullet & \bullet & \bullet & c \\
\bullet & \bullet & \bullet & d \end{array} \right), \: \: \:
S^{-1} \: = \: \left( \begin{array}{cccc}
\bullet & \bullet & \bullet & \bullet \\
\bullet & \bullet & \bullet & \bullet \\
\bullet & \bullet & \bullet & \bullet \\
\alpha & \beta & \gamma & \delta \end{array} \right) \]
(where $a, b, c, d, \alpha, \beta, \gamma, \delta$ are barred quaternion
numbers) we obtain from eq.(B.3)
$$ \left( \begin{array}{cccc}
a\alpha & a\beta & a\gamma & a\delta \\
b\alpha & b\beta & b\gamma & b\delta \\
c\alpha & c\beta & c\gamma & c\delta \\
d\alpha & d\beta & d\gamma & d\delta \end{array} \right) \: = \:
\left( \begin{array}{cc}
1 + \frac{1}{2} B^{2 \, new} & 0 \\
\bullet & 1 + \frac{1}{2} C^{2 \, new} \end{array} \right) \eqno(B.4) $$
(where $1+\frac{1}{2}B^{2 \, new}, \bullet, 1+\frac{1}{2}C^{2 \, new}$ are
quaternion $2\times2$ matrices). Thus, in particular
\[ \left( \begin{array}{cc}
a\gamma & a\delta \\
b\gamma & b\delta \end{array} \right) \: = \:
\left( \begin{array}{cc}
0 & 0 \\
0 & 0 \end{array} \right) \]
This last equation leads to a contradiction. From it, it follows necessarily
that at least one of the couples ($a, b$) and ($\gamma, \delta$) must
be null. As a consequence,
\[ (a, b) \, = \, (0, 0) \: \: \Rightarrow \: \: B^{2 \, new} \: = \:
- 2 \left( \begin{array}{cc} 1 & 0\\ 0 & 1 \end{array} \right) \]
\[ (\gamma, \delta) \, = \, (0, 0) \: \: \Rightarrow \: \: C^{2 \, new} \:
= \: - 2 \left( \begin{array}{cc} 1 & 0\\ 0 & 1 \end{array} \right) \]
and this is not possible because $B^{2 \, new}$ and $C^{2 \, new}$
represent spin 1 $\oplus$ 0 and thus they cannot be proportional to the
identity.

\pagebreak

\end{document}